\documentclass[10pt,twocolumn,letterpaper]{article}

\usepackage{cvpr}
\usepackage{times}
\usepackage{epsfig}
\usepackage{graphicx}
\usepackage{amsmath}
\usepackage{amssymb}

\usepackage{amsmath}
\usepackage{amssymb}
\usepackage{array,multirow}
\usepackage{floatrow}
\usepackage[breaklinks=true,bookmarks=false,pagebackref=true,breaklinks=true,letterpaper=true,colorlinks,bookmarks=false]{hyperref}
\usepackage{float}
\hypersetup{
    colorlinks=true,
    linkcolor=blue,
    filecolor=magenta,      
    urlcolor=cyan,
    pdftitle={arXiv:2108.06490},
    pdfpagemode=FullScreen,
    }

\usepackage[pagebackref=true,breaklinks=true,letterpaper=true,colorlinks,bookmarks=false]{hyperref}

\cvprfinalcopy % *** Uncomment this line for the final submission

\def\cvprPaperID{****} % *** Enter the CVPR Paper ID here

\ifcvprfinal\fi
\begin{document}

\title{DICOM Imaging Router: An Open Deep Learning Framework for Classification of Body Parts from DICOM X-ray Scans}

\author{Hieu H. Pham$^{1,2,\dag}$, Dung V. Do$^{1}$, Ha Q. Nguyen$^{1,2}$ \\
$^{1}$Medical Imaging Center, Vingroup Big Data Institute, Hanoi, Vietnam\\
$^{2}$College of Engineering \& Computer Science, VinUniversity, Hanoi, Vietnam\\
{  $\dag$Corresponding author \tt v.hieuph4@vinbigdata.org}
}

\maketitle
%\thispagestyle{empty}

%%%%%%%%% ABSTRACT
\begin{abstract}
    X-ray imaging in Digital Imaging and Communications in Medicine (DICOM) format is the most commonly used imaging modality in clinical practice, resulting in vast, non-normalized databases. This leads to an obstacle in deploying artificial intelligence (AI) solutions for analyzing medical images, which often requires identifying the right body part before feeding the image into a specified AI model. This challenge raises the need for an automated and efficient approach to classifying body parts from X-ray scans. Unfortunately, to the best of our knowledge, there is no open tool or framework for this task to date. To fill this lack, we introduce a DICOM Imaging Router that deploys deep convolutional neural networks (CNNs) for categorizing unknown DICOM X-ray images into five anatomical groups: abdominal, adult chest, pediatric chest, spine, and others. To this end, a large-scale X-ray dataset consisting of 16,093 images has been collected and manually classified. We then trained a set of state-of-the-art deep CNNs using a training set of 11,263 images. These networks were then evaluated on an independent test set of 2,419 images and showed superior performance in classifying the body parts. Specifically, our best performing model (i.e., MobileNet-V1) achieved a recall of 0.982 (95\% CI, 0.977--0.988), a precision of 0.985 (95\% CI, 0.975--0.989) and a F1-score of 0.981 (95\% CI, 0.976--0.987), whilst requiring less computation for inference (0.0295 second per image). Our external validity on 1,000 X-ray images shows the robustness of the proposed approach across hospitals. These remarkable performances indicate that deep CNNs can accurately and effectively differentiate human body parts from X-ray scans, thereby providing potential benefits for a wide range of applications in clinical settings. The dataset, codes, and trained deep learning models from this study will be made publicly available on our project website at \url{https://vindr.ai/datasets/bodypartxr}.
\end{abstract}

%%%%%%%%% BODY TEXT
\section{Introduction}

X-ray is the most commonly performed procedure in  clinical practice. More than 600 million X-ray examinations are conducted yearly~\cite{christiansen2005x} for evaluating various human body parts such as the lungs, heart size, bowel, and bones. In recent decades, many automatic medical image analysis systems, particularly deep learning-based systems,  have been studies and deployed  to support radiologists in interpreting X-ray scans. To date, hundred AI software products for clinical radiology~\cite{van2021artificial} have been introduced. These systems are often developed for analyzing specific anatomies (\textit{e.g.}, lung, abdominal, spine, etc.) and often require the identification of the human body contained in the input image. Vast, non-normalized databases of X-ray images from hospitals raise  the need  for  an  automated  approach  to  classify  body parts from  X-ray  scans. An automatic system for accurate classification of body parts from X-ray scans helps identify the right input for AI systems. It is also a useful tool for data management at hospitals or medical centers.  Several body part recognition systems, which were relied on carefully hand-crafted features, have been introduced~\cite{aboud2015automatic,jeanne2009automatic}. In particular, machine learning-based algorithms~\cite{aboud2015automatic,saha2016classifying} have been applied and shown their superior performance on this task. We observed two limitations of the existing approaches. First, these methods were developed and tested on ImageCLEF’s 2015 -- a quite small dataset with 500 training images and 250 test images. This fact raises concerns~\cite{oakden2020exploring} about the robustness of the predictive models in real clinical contexts. Second, an automatic body part recognition system plays as an image router that requires a near-perfect level of performance (100\%) in recognizing the images. Meanwhile, the existing approaches reported a performance of about 80\%--85\% in accuracy, which is not confident enough to deploy in real-world clinical settings. Hence, this work aims to develop a highly accurate deep learning-based system for grouping unknown X-ray images into five anatomical groups:  abdominal X-ray, adult chest X-ray, pediatric chest X-ray, spine X-ray, and others.  To this end, a large-scale X-ray dataset consisting of 16,093 images has been collected and manually classified. We then trained a set of state-of-the-art deep CNNs using a training set of 11,263 images. These networks were then evaluated on an independent test set of 2,419 images and showed superior performance in classifying the body parts while requiring less computation for inference. To summarize, the main contributions of this work two folds:
\begin{itemize}
    \item We introduce and release a large-scale dataset for the classification of body parts from X-ray scans. The dataset contains 16,093 X-ray images in DICOM format, for which each was manually annotated for five anatomical groups: abdominal X-ray, adult chest X-ray, pediatric chest X-ray, spine X-ray, and others. To the best of our knowledge, this is the largest X-ray dataset for human body part classification task to date. It will be opened for public access from \url{https://vindr.ai/datasets/bodypartxr}. 
    
    \item We develop a robust DICOM Imaging Router that used a state-of-the-art deep CNN model to classify X-ray images based on the presence of the body part in the image. Our experimental results show superior performance on an independent test set while requiring less computation for inference. The proposed system potential benefits for a wide range of applications in clinical settings. It was made publicly available at\footnote{\url{https://github.com/vinbigdata-medical/DICOM-Imaging-Router}} for the community as an open deep learning framework that can be easily reused and finetuned. 

\end{itemize}

\begin{figure*}
    \centering
    \includegraphics[width=13cm,height=5.5cm]{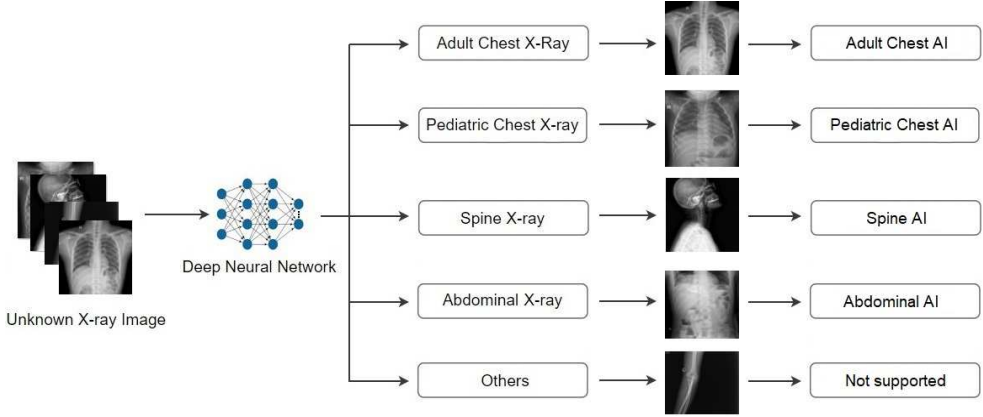}
    \caption{We develop a deep learning-based classifier for automatic recognition of body parts from X-ray scans. Given an unknown X-ray as input, the system is able to classify the scan into one of five groups, including adult chest X-ray, pediatric chest X-ray, spine X-ray, abdominal X-ray, and others. In a simple practical scenario, each classified image can be then passed through the corresponding AI model. }
    \label{fig:system}
\end{figure*}

\section{Methodology}

\subsection{DICOM Imaging Router: System overview}

An overview of the DICOM Imaging Router is illustrated in Figure~\ref{fig:system}.  It is a deep learning-based classifier that accepts an unknown X-ray as input and classifies it into one of five groups, including abdominal X-ray, adult chest X-ray, pediatric chest X-ray, spine X-ray, and others. From a practical point of view, a reliable DICOM Image Router should ensure two essential requirements, including (1) a nearly 100\% classification accuracy, and (2) a low inference time. To achieve these goals, we collect and annotate a large-scale X-ray dataset. We then train a set of state-of-the-art lightweight CNN models. Mathematically, this is a supervised multi-class classification task task that assigns a class label for each input example. Given a training dataset of $N$ labeled examples of the form $\left\{\left(\textbf{x}^{(i)}, y^{(i)}\right)\right\}$, where $\textbf{x}^{(i)} \in \mathbb{R}^{n}$ is the \textit{i}-th X-ray example and $y^{(i)} \in {1, . . . , K}$ is the \textit{i}-th class label. Here, $K$ denotes the number of classes. In this task, we aim at building a
learning model $f_{\boldsymbol{\theta}}$ such that it classifies accurate for new unseen examples~\cite{aly2005survey}. This task can be done by training a deep CNN that learns a non-linear mapping from the input $\textbf{x}^{(i)} \in \mathbb{R}^{n}$ to the corresponding label 
$y^{(i)} = f_{\boldsymbol{\theta}}(\textbf{x}^{(i)}) \in \mathbb{R}^{K}$.  One common solution to train the network is to minimize the softmax cross-entropy loss
\begin{align}
\mathcal{L}(\boldsymbol{\theta}) = - \sum_{i=1}^{N} y^{(i)} \log (\sigma(f_{\boldsymbol{\theta}}(\textbf{x}^{(i)}))) 
\end{align}
over all $N$ training examples. Here the standard softmax function $\sigma : \mathbb{R}^K \to [0,1]^{K}$ is defined by the formula
\begin{align}
\sigma(\textbf{z})_i = \frac{e^{(z_i)}}{\sum_{j=1}^{K}e^{(z_j)}}
\end{align} 
for $i = 1, ..., K$ and $\textbf{z} = (z_1, ...z_K) \in \mathbb{R}^{K}$. 
\subsection{Data collection and annotation}
The dataset used in the study was collected from the Picture Archiving and Communication System (PACS) of several major hospitals. The ethical clearance of this study approved by the IRB of each hospital before any research activities. All patient-identifiable information in the data has been removed. The need for obtaining informed patient consent was waived because this study did not impact clinical care or workflow at the hospital. We recruited a group of human readers to participate in our labeling labeling process. Specifically, all X-ray scans were manually reviewed and classified case-by-case into five groups: abdominal X-ray, adult chest X-ray,  pediatric chest X-ray,  spine X-ray,  and others. In particular, each example was manually classified into two rounds by two different readers. In total, 16,093 images have been collected and manually categorized. We used a stratified random sampling method for dividing the dataset into train, validation, and test set with respective ratios of 0.7/0.15/0.15. As a result, 11,263 images will be used to train deep learning algorithms, 2,411 and 2,419 images will be used as validation and test sets, respectively, for evaluating the algorithms. Each image was then stored in the .PNG format and rescaled to the size of $512\times512$ pixels. Table~\ref{tab:dataset} below summarizes the data sets used in this study.
\begin{table}[H]
\centering
{\caption{Details of training, validation, and test data sets used in this study.  To the best of our knowledge, this is the largest X-ray dataset for human body part classification tasks to date.} \label{tab:dataset}}
\scriptsize{
\begin{tabular}{|c|c|c|c|c|}
\hline
\textbf{Body part} & \textbf{Training set} & \textbf{Validation set} & \textbf{Test set} & \textbf{Total} \\\hline
Abdominal X-ray & 825 & 176 & 178 & 1,179 \\
Adult chest X-ray & 2,304 & 493 & 495 & 3,292 \\
Pediatric chest X-ray & 4,352 & 932 & 934 & 6,218 \\
Spine X-ray & 1,559 & 334 & 335 & 2,228 \\
Others & 2,223 & 476 & 477 & 3,176 \\
\hline
\textbf{All categories} & \textbf{11,263} & \textbf{2,411} & \textbf{2,419} & \textbf{16,093} \\
\hline
\end{tabular}}
\end{table}

\begin{table*}
\small{
\centering
{\caption{Classification performance of different network architectures on the test set. Inference time (in second) is measured on an RTX 2080 Ti GPU machine. Best results are in \textcolor{red}{\textbf{red}}.} \label{tab:result}}%  
{\begin{tabular}{|c|c|c|c|c|c|}
\hline
 \textbf{Model} & \textbf{Recall}  &  \textbf{Precision} &  \textbf{\textit{F1}-scor}e &  \textbf{Inference Time} &  \textbf{\# Parameters}\\
\hline
\textcolor{red}{\textbf{MobileNet-V1}~\cite{howard2017mobilenets}} & \textcolor{red}{\textbf{0.982 (0.977--0.988)}} & \textcolor{red}{\textbf{0.981 (0.975--0.987)}} & \textcolor{red}{\textbf{0.981 (0.976--0.987)}} & \textcolor{red}{\textbf{0.0295}} & \textcolor{red}{\textbf{3,2M}}\\
\hline
MobileNet-V2~\cite{sandler2018mobilenetv2} & 0.967 (0.985--0.976) &  0.979 (9.974--0.985) & 0.972 (0.965--0.980) & 0.0322 & 3,5M\\
\hline
ResNet-18~\cite{he2015deep} & 0.923 (0.909--0.937) &  0.939 (0.927--0.951) &  0.930 (0.917--0.942) & 0.0324 & 11,6M\\
\hline
ResNet-34~\cite{he2015deep} & 0.923 (0.909--0.937) & 0.935 (0.923--0.948) & 0.929 (0.916--0.941) & 0.0350 & 21,7M\\
\hline
EfficientNet-B0~\cite{tan2019efficientnet} & 0.975 (0.968--0.981) & 0.980 (0.975--0.986) & 0.977 (0.971--0.983) & 0.0352 & 14,1M\\
\hline
EfficientNet-B1~\cite{tan2019efficientnet} & 0.969 (0.961--0.977) & 0.977 (0.971--0.983) & 0.973 (0.966--0.980) & 0.0381 & 27,2M\\
\hline
EfficientNet-B2~\cite{tan2019efficientnet} & 0.973 (0.965--0.980) & 0.977 (0.972--0.984) &  0.975 (0.969--0.982) & 0.0384 & 29,4M\\
\hline
\end{tabular}}}
\end{table*}

\subsection{Deep learning algorithms}

To classify body parts from X-ray images, we exploited state-of-the-art, light-weight CNNs that have achieved remarkable performance on many image classification tasks, including MobileNet-V1~\cite{howard2017mobilenets}, MobileNet-V2~\cite{sandler2018mobilenetv2}, ResNet-18~\cite{he2015deep}, ResNet-34~\cite{he2015deep}, and EfficientNet-B0/B1/B2~\cite{tan2019efficientnet}. We followed the original implementations~\cite{howard2017mobilenets,sandler2018mobilenetv2,he2015deep,tan2019efficientnet} with minor modifications. Specifically, we replaced the last fully connected layer of each architecture with a new layer of 5 neurons, corresponding to the number of body parts. During the training stage, we rescaled all training images to 512$\times$512. All models were trained using cross-entropy loss function with Adam optimizer~\cite{kingma2014adam}. The learning rate was set at $1\times e^{-4}$ and then simulated warm restarts by scheduling the learning rate~\cite{loshchilov2016sgdr}. All networks were trained for 100 epochs using Pytorch (v1.7.0) on a machine with one RTX 2080 Ti GPU.

\section{Experiments and Results}
\subsection{Experimental setup and evaluation metrics}
We evaluated the performance of the proposed models on an internal test set (\textit{N = 2,419}) and an external (\textit{N = 1,000}) test set  using precision, recall, \textit{F1}-score and mean inference time (in second on GPU) per image. Using the final prediction provided by the models and the ground truth labels, we calculated the true positives (TPs), true negatives (TNs), false positives (FPs), and false negatives (FNs) as Table~\ref{tab:confusion_matrix}.
\begin{table}[H]
\centering
{\caption{Confusion matrix} \label{tab:confusion_matrix}}
\small{
\begin{tabular}{c|c|c}
\textbf{} & \textbf{Actually positive} & \textbf{Actually negative} \\
\hline
\textbf{Predicted
positive}& TPs & FPs  \\
\hline
\textbf{Predicted
positive} & FNs & TNs \\
\end{tabular}}
\end{table}
\noindent  The precision, recall and \textit{F1}-score were then computed by
\begin{align}
\text{precision} = \frac{\text{TPs}}{\text{TPs + FPs}},
\end{align}
\begin{align}
\text{recall} =\frac{\text{TPs}}{\text{TPs+FNs}},
\end{align}
\begin{align}
\text{\textit{F1}-score} = \frac{2}{\text{precision}^{-1}+\text{recall}^{-1}}.
\end{align}
For each measure, we estimated 95\% bootstrap confidence interval with 10,000 iterations. 

\subsection{Model performance on internal test set}

Table~\ref{tab:result} summarizes quantitative results for all the classification models. Deep CNNs showed excellent performances on 2,419 of the external test set. Specifically, our best performing model (\textit{i.e.}   MobileNet-V1~\cite{howard2017mobilenets}, 3.2M) achieved a recall of 0.982 (95\% CI, 0.977--0.988), a precision of 0.981 (5\% CI, 0.975--0.987) and a \textit{F1}-score of 0.981 (95\% CI, 0.976--0.987), whilst requiring less computation for inference (0.0295 second per image). 

\subsection{Model performance on external test set}

The domain shift across different hospital settings is the main obstacle in transferring deep learning models into clinical practice~\cite{pooch2019can}. It can result in poor generalization and decreased accuracy~\cite{guan2021domain}. To investigate the generalization ability of the proposed approach across multiple data sources, we performed an external validation test on 1,000 X-ray images collected from another patient cohort. The best-performing model MobileNet-V1~\cite{howard2017mobilenets} was used for this experiment. It reported a recall of 0.9712, a precision of 0.9738, and an \textit{F1}-score of 0.9725. This high diagnostic accuracy shows the robustness of the system across different patient cohorts, scanner vendors, and imaging protocols without additional training cost.

\section{Conclusions}

This work developed and validated a deep learning-based DICOM Imaging Router to classify body parts from X-ray images.  A benchmark dataset with 16,093 X-ray images of body parts has been introduced. Experiments demonstrated the  effectiveness of the proposed method. The DICOM Imaging Router can be applied for many real-world applications in radiology. For example, it can be integrated into a PACS system to help radiologists find and classify X-ray images quickly and accurately for interpretation. The system can play the role of pre-filter for other AI applications. Our trained models and dataset used in this study will be opened for further development and deployment. For future work, we plan to conduct more experiments and evaluate the impact of the proposed framework in real-world clinical settings.

\bibliographystyle{ieee}
\bibliography{egbib}

\end{document}